# Direct probe of anisotropy in atom-molecule collisions via quantum scattering resonances

Ayelet Klein[1*], Yuval Shagam[1*], Wojciech Skomorowski[2], Piotr S. Żuchowski[3], Mariusz Pawlak[4], Liesbeth M. C. Janssen[5], Nimrod Moiseyev[6], Sebastiaan Y. T. van de Meerakker[7], Ad van der Avoird[7], Christiane P. Koch[2] and Edvardas Narevicius[1]

**Anisotropy is a fundamental property of particle interactions. It occupies a central role in cold and ultra-cold molecular processes, where long-range forces have been found to significantly depend on orientation in ultra-cold polar molecule collisions[1,2]. Recent experiments have demonstrated the emergence of quantum phenomena such as scattering resonances in the cold collisions regime due to quantization of the intermolecular degrees of freedom[3–8]. Although these states have been shown to be sensitive to interaction details, the effect of anisotropy on quantum resonances has eluded experimental observation so far. Here, we directly measure the anisotropy in atom-molecule interactions via quantum resonances by changing the quantum state of the internal molecular rotor. We observe that a quantum scattering resonance at a collision energy of $k_B \; x \; 270 \; mK$ appears in the Penning ionization of molecular hydrogen with metastable helium only if the molecule is rotationally excited. We use state of the art *ab initio* and multichannel quantum molecular dynamics calculations to show that the anisotropy contributes to the effective interaction only for $H_2$ molecules in the first excited rotational state, whereas rotationally ground state $H_2$ interacts purely isotropically with metastable helium. Control over the quantum state of the internal molecular rotation allows us to switch the anisotropy on or off and thus disentangle the isotropic and anisotropic parts of the interaction. These quantum phenomena provide a challenging benchmark for even the most advanced theoretical descriptions, highlighting the advantage of using cold collisions to advance the microscopic understanding of particle interactions.**

[1]Department of Chemical Physics, Weizmann Institute of Science, Rehovot 76100, Israel.
[2]Theoretische Physik, Universität Kassel, Heinrich-Plett-Straße 40, 34132 Kassel, Germany.
[3]Institute of Physics, Faculty of Physics, Astronomy and Informatics, Nicolaus Copernicus University in Torun, Grudziadzka 5, 87-100 Torun, Poland.
[4]Faculty of Chemistry, Nicolaus Copernicus University in Toruń, Gagarina 7, 87-100 Toruń, Poland.
[5]Institute for Theoretical Physics II, Heinrich-Heine University, Universitaetsstrasse 1, D-40225 Duesseldorf, Germany.
[6]Schulich Faculty of Chemistry and Faculty of Physics, Technion-Israel Institute of Technology, Haifa 32000, Israel.
[7]Radboud University, Institute for Molecules and Materials, Heijendaalseweg 135, 6525 AJ Nijmegen, Netherlands.
*These authors contributed equally to this work.

Anisotropy leaves fingerprints on many atomic and molecular processes, but its contribution is difficult to evaluate and quantify at high energies due to the many quantum states needed to describe the process. For example, anisotropy effects can be traced to changes in the angular distribution of products in collisions and half-collisions that are detected with advanced ion imaging techniques[9]. Numerous experiments have reported orientation effects including the study of bimolecular processes such as the reactive scattering of $H + D_2$ ref. (10), and $F + CD_4$ ref. (11) as well as the inelastic scattering of $NO$ by $Ar$ ref. (12). Spectroscopic techniques have been successfully used to identify hindered $H_2$ rotation in an anisotropic potential of van der Waals molecules such as $H_2O + H_2$ complexes[13] and $H_2HF$ complexes[14].

In cold and ultra-cold molecular processes, where the number of accessible quantum states is greatly reduced, anisotropy effects have so far been studied only in fast reactions. It has been shown that orientation of the long-range dipole-dipole interaction between molecules can be tuned via external electric fields[15]. In $KRb$ molecules this method enabled the observation of rethermalization[1] and the suppression of a bimolecular chemical reaction[2]. In addition, the reaction of $N_2^+ + Rb$ has been shown to be dominated by the anisotropic charge-quadrupole long range interaction[16]. Similarly, fast Penning ionization reactions have been studied, demonstrating that the rotational state of a molecule defines the type and strength of the effective long-range interaction[17]. These fast processes are fully described by the long-range interactions alone. At short internuclear separations the Penning ionization happens with unit probability acting as a perfect absorber preventing the formation of outgoing waves. In such a case there is no quantization along the internuclear coordinate which precludes the appearance of quantum scattering resonances.

Experiments that are capable of resolving quantum resonance states allow for collision studies with near spectroscopic precision. For example, isotope shifts have been observed in the Penning ionization of molecular hydrogen isotopologues[6]. Similar effects between isotopes were observed in rotational state-to-state cross sections and linked to quantum scattering resonances appearing at different energies[18]. Feshbach resonances in ultra-cold anisotropic atom-atom collisions allowed the emergence of chaotic scattering in ultra-cold $Er$ and $Dy$ to be observed[19].

The emergence of discrete quantized metastable states formed during a cold collision opens a new way to study anisotropy. Instead of recording the angular distribution of the collision products, one can conveniently alter the way anisotropy contributes by choosing different rotational quantum states of the interacting molecules. The sensitivity of orbiting resonances towards molecular degrees of freedom has been theoretically predicted for example for the reaction $D + H_2$ ref. (20), the reactive scattering of $F + H_2$ ref. (21), and inelastic collisions of $CO + H_2$ ref. (22). This effect has so far eluded experimental investigation due to insufficient resolution of individual quantum resonance states[22]. The introduction of merged beam experiments reaching temperatures as low as 10 mK[3,6] allow for both low enough collision energies and high resolution, due to the correlation that develops during the free propagation of a supersonic molecular beam[23]. This merged beam configuration is used in the current collision measurements, as described in the methods section.

Here we demonstrate the dramatic effect of anisotropy on shape resonances in cold Penning ionization reactions of metastable helium with molecular hydrogen in the rotational ground and first excited states (Fig. 1). In the classical collision regime down to several Kelvin temperatures, $H_2$ behaves similarly, independent of whether it is in the $j = 0$ or $j = 1$ internal rotational state. However, below one Kelvin we observe a striking difference. While a strong shape resonance exists for $j = 1$ at a collision energy of $k_B \times 270\ mK$, no resonances are observed in this energy range in the case of $j = 0$. Using state of the art *ab-initio* theory and coupled channels quantum molecular dynamics we find that the interaction between hydrogen, which is internally in the ground rotational state and metastable helium, is effectively isotropic. This holds for any collision where the interaction anisotropy is small compared to the rotational constant of the molecule. In such a case, the anisotropic part of the interaction potential becomes important only for molecules in an excited rotational state. The wavefunction of rotationally excited $H_2$ molecules is not spherically symmetric causing anisotropic components of the interaction to come into play, while in the rotational ground state, where the wavefunction is spherically symmetric, the contribution of anisotropic components is zero. We show that by accurately measuring the orbiting resonance positions, we can independently probe the isotropic and anisotropic parts of the interaction. Since the resonance wavefunctions span both the short and the long range parts of the potentials, we are able to benchmark first principles quantum dynamics calculations to within $7 \cdot 10^{-3}\ cm^{-1}$ or 200 MHz precision. This allows for a rigorous comparison with theory and a quantitative evaluation of different levels of treating electron correlations in electronic structure calculations.

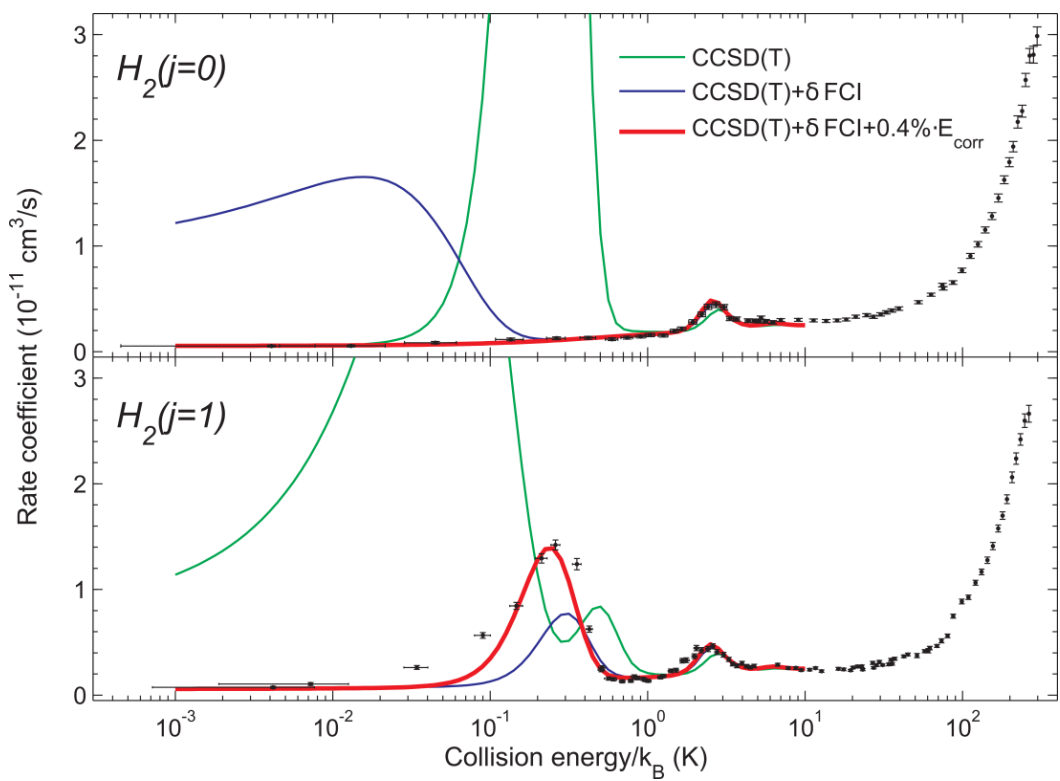

**Figure 1.** The Penning ionization rate coefficient as function of collision energy of $He^*(2^3S_1) + H_2$ from ~$300K$ down to ~$10\ mK$ is presented in black dots including error bars. The upper (lower) panel corresponds to $H_2$ internally in the $j = 0$ ($j = 1$) rotational quantum state. Two shape resonances are observed below $5\ K$ for the reaction with ortho-$H_2$ ($j = 1$) at $2.37\ K$ and at $270\ mK$. The $j = 1$ lower energy resonance state corresponds asymptotically to the partial wave $l = 3$, while the second resonance appearing both for $j = 0$ and $j = 1$ correspond asymptotically to the $l = 4$ partial wave. In the reaction with para-$H_2$ ($j = 0$), the $l = 3$ resonance is absent, while the difference in energy of the $l = 4$ resonance cannot be distinguished from the $j = 1$ case. The results of state-of-the-art first principles calculations for the reaction with the ground ($j = 0$) and excited ($j = 1$) rotational states are depicted by the green, blue, and red curves in both panels. The interaction potential obtained with the current "gold standard" of electronic structure methods, the coupled cluster theory with singles, doubles and non-iterative triple excitations (CCSD(T), green), alone erroneously predicts a low energy resonance for para-hydrogen ($j = 0$) and two low energy resonances for ortho-hydrogen ($j = 1$). Including the full configuration interaction (FCI) correction (blue) allows for agreement down to collision energies corresponding to a few hundred milli-Kelvin. Further improvement of the interaction potential is achieved by uniformly scaling the correlation energy by 0.4% (red), resulting in excellent agreement with the measured resonance positions and the overall behavior of the rate coefficient down to the lowest collision energies probed in the experiment.

Conveniently, merged beam experiments with molecular hydrogen can be carried out with control over the rotational state of the molecule. A normal molecular hydrogen beam cooled in supersonic expansions consists of 75% ortho-hydrogen and 25% para-hydrogen. It follows from the Pauli principle that the lowest rotational state for para-hydrogen is $j = 0$ whereas $j = 1$ is the lowest molecular rotational state for ortho-hydrogen. A pure para-hydrogen sample can be prepared by catalytic conversion at cryogenic temperatures allowing the determination of the state selected rate coefficient, as has been described previously[17].

In Fig. 1, we compare our experimental data to the results of first principles calculations. A 2D potential energy surface was obtained with the current "gold standard" of electronic structure theory, the coupled cluster method with single, double and approximate noniterative triple excitations (CCSD(T)), and further refined to include higher order correlation effects. Subsequently, the potential energy surface was used in quantum scattering calculations with reactive boundary conditions at short range[24]. The theoretical results demonstrate that indeed collisions with $j = 0$ hydrogen are not sensitive to the anisotropic part of the interaction potential, such that the hydrogen molecule effectively behaves like an atom. Even scaling the anisotropic part by 50% does not introduce an observable change in the rate coefficient, as can be seen in the inset in Fig 2. In stark contrast the $j = 1$ hydrogen rate coefficient is very sensitive to the anisotropy, emerging as shifts in the position of the orbiting resonance for different scaling factors of the anisotropic potential shown in Fig 2.

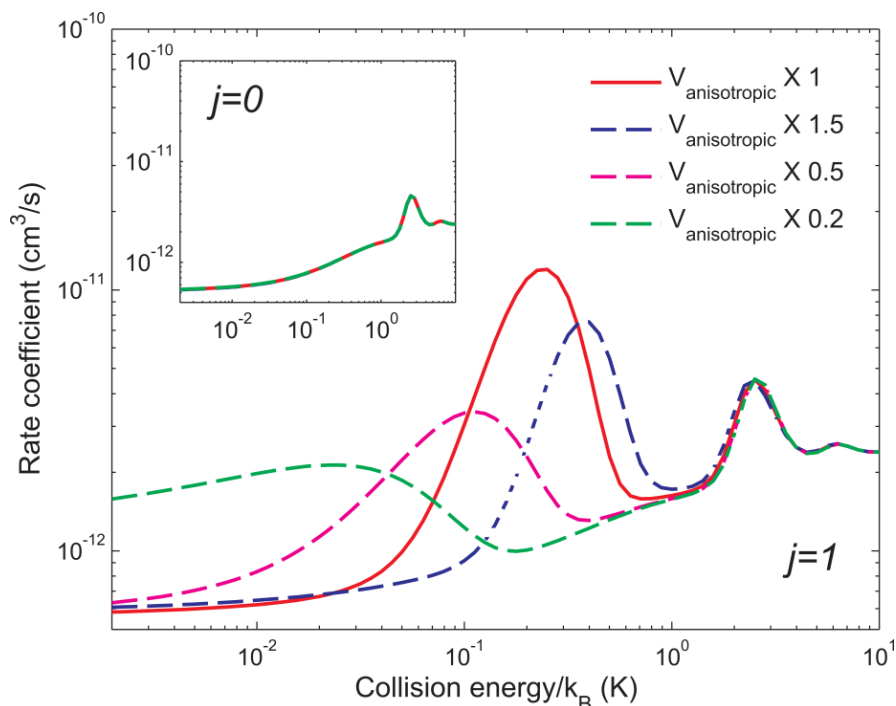

**Figure 2.** The theoretical rate coefficients of $He^*(2^3S) + H_2$ in the first excited rotational state ($j = 1$) for different scaling factors of the anisotropic potential $V_2(R)$. The rate resulting from the original potential is depicted by the solid red curve, and the scaling of the anisotropic part by a factor of 1.5, 0.5 and 0.2 is depicted by the blue, magenta and green dashed curves respectively. As the scaling factor decreases the lower energy resonance position shifts to lower collision energies. The spectroscopic resolution of the resonance positions can serve as a sensitive measure of the anisotropy in the system. The logarithmic inset shows the rate for scaling factors of the anisotropic potential by the same factors for the reaction of $He^* + H_2$ in the ground rotational state ($j = 0$). The rate coefficient remains the same for different scaling factors, since the anisotropic potential has limited contribution to this reaction.

The sensitivity of quantum resonance states towards small changes in the interaction allows us to make a meaningful comparison of different levels of treating electron correlations in *ab initio* theory (Fig. 1). In fact, the interaction potential obtained with the "bare" CCSD(T)/aug-cc-pV6Z method is not accurate enough to reproduce the low-energy resonances for collisions with both $j = 0$ and $j = 1$ $H_2$ (see the green curves in Fig. 1). To account for higher order electron correlation effects, the full configuration interaction (FCI) contribution beyond CCSD(T) has been calculated and added to the potential. The FCI-corrected potential energy surface performs significantly better (blue curves in Fig. 1), in particularly for collisions with $H_2(j = 1)$. However, it still predicts a fictitious resonance in the milli-Kelvin range for $H_2(j = 0)$. To further improve the theoretical model, we have uniformly scaled the correlation energy. This is motivated by the fact that incompleteness of the orbital basis set is the main source of error in the *ab initio* calculations. It particularly affects the correlation energy whereas the Hartree-Fock energy is practically fully converged in the aug-cc-pV6Z basis set (see Methods for details). Indeed, increasing the correlation energy in the interaction potential by merely 0.4% leads to an excellent agreement for the resonance position of $H_2(j = 1)$ and confirms absence of this resonance for $H_2(j = 0)$ (red curves in Fig. 1). The latter can be understood as follows: Scaling the correlation energy by 0.4% corresponds to increasing the depth of the isotropic potential by around 0.1 cm$^{-1}$. This shifts down the erroneous resonance obtained with CCSD(T) and CCSD(T) + δ FCI such that it becomes a truly bound state and does not anymore affect the reaction rate.

A similar comparison of different levels of theory and experiment could in principle also be performed using rovibrational spectroscopy of bound states. However, while spectroscopic methods measure the spacing between rovibrational states with high precision, they are susceptible to global shifts of the spectrum relative to the dissociation limit. In collisional spectroscopy the spectrum emerges above the dissociation threshold enabling the measurement of the energies of quasi-bound states on an absolute scale. Thus changes in the effective well depth due to the anisotropy are more pronounced in absolute scale measurements of the spectrum (Fig. 3).

In order to qualitatively understand the role of anisotropy, it is convenient to consider the effective adiabatic interaction potentials. To this end, the interaction potential is expanded into two terms, $V_0(R)$ the isotropic term, where $R$ is the interparticle separation, and $V_2(R)P_2(\cos\theta)$ which describes the angular dependence to first order, where $\theta$ is the polar angle between the $H_2$ molecule axis and the axis joining $H_2$ center of mass with the $He$ atom. We use an adiabatic theory[25] inspired by Klemperer and colleagues[26] as well as Scribano et al.[27] in order to construct effective potential curves that conserve the total angular momentum $J$. These potentials asymptotically correlate to a well-defined rotational state of $H_2$ ($j$) and a well-defined collisional angular momentum, i.e., partial wave $l$. The anisotropy, $V_2(R)$, is small in our case compared to the rotational constant for all collision distances, such that mixing of rotational states of the same parity in the $H_2$ molecule is negligible. Figure 3 compares the three most relevant adiabatic potentials for the reaction that asymptotically correlate with $j = 0$ and $j = 1$ $H_2$. The red and black plotted adiabats correspond to scattering channels with the total angular momentum $J = 3$ and orbital angular momentum (partial wave) $l = 3$, i.e. the channel for which the resonance occurs. The black curve corresponding to $J = 3$ and $j = 0$ $H_2$ is identical to the isotropic part of the interaction potential and includes the centrifugal term. It is deep enough to support a single bound state just below the threshold with a binding energy of $k_B$ $x$ $0.01$ $K$ as shown in the inset of Fig. 3. In case of $j = 1$ $H_2$, the anisotropy is responsible for reducing the well depth of the corresponding $J = 3$ effective potential, shown in red, enough to raise the bound state above the dissociation threshold (Fig. 3 inset). This state is located below the top of the centrifugal barrier, and is accessible via tunneling in scattering experiments at a collision energy of $k_B$ $x$ $270$ $mK$. The grey adiabat in Fig. 3 represents two nearly degenerate effective potential curves with the total angular momentum of $J = 2$ and $J = 4$ and $j = 1$ $H_2$. Both potentials do not support resonances and have bound levels at $k_B$ $x$ $0.18$ $K$ below the dissociation threshold.

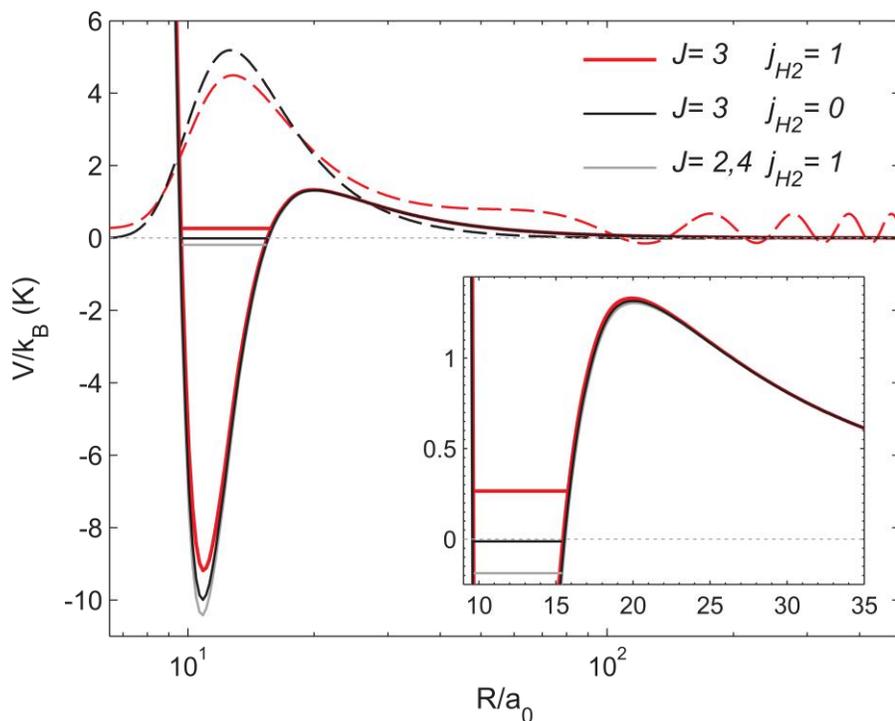

**Figure 3.** Effective potential curves for the reaction of metastable $He$ and $H_2$, internally in the ground (solid black curve) and excited (solid red and grey curves) rotational states are presented, calculated according to adiabatic theory[25]. The adiabatic potentials shown in red and black correspond to the total angular momentum $J = 3$. The potential of $j = 0$ supports one bound state, located below the dissociation threshold at $k_B \, x \, 0.01 \, K$. The contribution of the anisotropic part of the potential in the case of $j = 1$ (red curve) causes a reduction of the well depth. This reduction is sufficient to shift the bound state by $\sim 0.3 \, K$ placing it above the dissociation threshold, as depicted by the horizontal solid red line, leading to the resonance at 270 mK in the reaction rate coefficient of $He^* + H_2$ in the excited rotational state ($j = 1$). The grey curve correspond to two nearly degenerate effective potential curves with total angular momentum of $J = 2$ and $J = 4$ and with $j = 1$ $H_2$. They support bound states at $k_B \, x \, 0.18 \, K$ below the dissociation threshold and no quasi-bound states. The bound and quasi-bound states wavefunctions for total angular momentum $J = 3$ are depicted by the black and red dashed curves respectively. The resonance state wavefunction spans both the short and the long range part of the potential, and can therefore be probed in collision experiments.

Orbiting resonances amplify minute differences in particle interactions, making them easily detectable in cold collision experiments, provided the energy resolution is sufficiently high. They are a manifestation of the quantization of intermolecular degrees of freedom forms a wave matter analogue of an optical cavity. Quantum resonances thus provide an ideal probe to illuminate different contributions to particle interactions, with a sensitivity that is a challenge for current state of the art *ab initio* theory. The sensitivity of quantum resonances towards the internal state of a molecule can be exploited in many experiments. Changing the rotational state of a molecule effectively switches the anisotropy of the interparticle interaction on and off, shifts resonance positions and thus dramatically changes the collisional cross section. Such control over collisional properties can be critical to the success of molecular cooling methods that rely on collisions such as evaporative or sympathetic cooling.

**Methods**

Methods and any associated references are available in the online version of the paper.

**Acknowledgements**

This research was made possible, in part, by the historic generosity of the Harold Perlman family. The authors acknowledge financial support from the European Commission through ERC grant EU-FP7-ERC-CoG 1485 QuCC. Additional financial support from the German-Israeli

Foundation, (grant no. 1254), from the COST Action (CM1405), as well as the Polish National Science Center grant DEC-2012/07/B/ST2/00235 is acknowledged.

**Competing financial interests**

The authors declare no competing financial interests.

**Methods**

The experimental setup, which has been discussed previously[3,6,17], consists of two supersonic beams, generated by pulsed Even-Lavie valves[28], placed at a relative angle of 10°. The straight beam consists of noble gas mixtures of $H_2$. The second beam consists of $^4He$, excited to the $2^3S_1$ metastable state ($He^*$) using a dielectric barrier discharge[29], located at the exit of the valve, and is merged with the straight beam using a $20\ cm$ magnetic guide, leading to zero relative angle between the beams at the detection region. The $He^*$ beam velocity is kept constant at $870\ \frac{m}{s}$ while the $H_2$ beam velocity is tuned by changing the valve temperature and seeding the $H_2$ in noble carrier gas mixtures. The collision energy is determined by the relative mean velocity between the beams and the velocity distribution of each beam spanning from $300\ K$ to $10\ mK$[23].

The metastable $He$ beam is characterized using an on-axis multichannel plate (MCP). The $H_2$ beam is characterized using a time-of-flight mass spectrometer (TOF-MS)[30], which contains an additional ionization element. The TOF-MS is positioned perpendicular to the beams' propagation, operating in a multi-pulsed mode, at $4\ \mu s$ intervals. The reaction product ions are detected with the TOF-MS with the ionization element turned off, while the neutral molecules are characterized with the ionization element turned on. The relative rate coefficient is calculated according to the ratio between the products' measured signal and the reactants' measured signal at each time interval. The rate coefficient is normalized to an absolute scale according to room temperature thermal rate measurement of $He(2^3S_1) + H_2$ ref. (31).

The para-$H_2$ sample is prepared by liquefying normal-$H_2$ in liquid Helium while in contact with the paramagnetic catalyst nickel sulfate. The sample is examined by resonance-enhanced multiphoton ionization (REMPI) followed by detection with the TOF-MS to determine its purity, showing that more than 98% of the $H_2$ is in the ground rotational state, as previously described in detail[32].

Rate coefficients were obtained by carrying out coupled-channels quantum reactive scattering calculations using the method by Janssen et al.[24], where only the reactant configuration is explicitly considered, and knowledge of the real value of interaction energy is sufficient. This approach is based on the assumption that the reaction proceeds irreversibly once the reactants reach a sufficiently short interparticle distance $R_c$, the "capture radius"; and reactive collisions are simulated by imposing appropriate short-range boundary conditions. Such a quantum capture approach is particularly suitable to describe a chemi-ionization process because of its irreversibility and strongly exoenergetic character, which implies that the total rate coefficient is exclusively determined by the dynamics in the entrance channel of the reaction. Since the real-valued Born-Oppenheimer interaction potential between $He^*$ and $H_2$ does not include any coupling with the ionization continuum, $He + H_2^+ + e^-$, we modified the purely repulsive wall of the potential, assuming that the isotropic part $V_0(R)$ is constant at short intermolecular distance ($R < R_c$) and equal to the energy released in the ionization process, i.e., $V_0(R) = V_c = -3.646\ eV$ if $R \leq R_c$. This results in a barrier at short range and allows for the loss of the flux into the reactive channels via quantum tunneling.

Close-coupling quantum scattering calculations were performed separately for collisions of $He^*$ with para-$H_2$ ($j = 0$) and with ortho-$H_2$ ($j = 1$). The angular basis functions were constructed as products of the rotational state of the $H_2$ molecule and the partial wave describing the orbital motion of the whole complex, and were symmetry-adapted to values of the total angular momentum $J$ and parity. The basis set was restricted to even or odd rotational states of $H_2$ for collisions with para-$H_2$, or ortho-$H_2$ respectively. For collision energies up to 10 Kelvin, the basis set yielding fully converged cross sections included all relevant functions with $j \leq 4$ and $J < 8$. The scattering wave functions were propagated from $R_c = 7.0$ to $R_{max} = 200\ a_0$ by means of a Numerov propagator. The capture radius was determined by the condition that the elastic cross sections are fully converged; choosing a smaller $R_c$ only results in a homogeneous scaling of the reaction rate coefficients, as does a modification of $V_c$. Short-range and asymptotic boundary conditions were imposed afterwards, as described in the original paper[24]. The calculated reactive cross sections as a function of energy were multiplied by the velocity and convoluted with the experimental uncertainty for the collision energies to yield the final theoretical rate coefficients.

The potential energy surface was obtained using the counter-poise corrected supermolecular method in which the interaction energy is obtained as the difference of the total energy of the dimer ($He^* + H_2$) and the total energies of the monomers, calculated in the basis set of the dimer with basis functions placed at the position of the interacting partner with charges set to zero. Since the $He^* + H_2$ system is not the actual triplet ground state of the system (which is $He(1^1S) + {}^3\Sigma$ excited $H_2$) we have employed the maximum overlap method[33,34] with the orbitals corresponding to the isolated $He^* + H_2$ potential with appropriate occupancy of $1s^1 2s^1 \sigma g^2$ as starting point. The spin-restricted coupled-cluster method with single- and double excitations with non-iterative correction to triple excitations (CCSD(T)) in a very large basis set, considered nowadays the gold standard of quantum chemistry, was employed, using the MOLPRO suite of codes[35]. The aug-mcc-pVXZ family for the hydrogen atom[36] and the aug-cc-pVXZ family optimized for the triplet state of the metastable helium dimer[37] were employed, where X=4,5,6,7 stands for the cardinal number, which corresponds to the highest angular momentum, equal to *l*=3,4,5,6, respectively. To better account for the dispersion interaction we have also used a set of mid-bond functions located between the center-of-mass of the $H_2$ molecule and the $He$ atom. These were obtained from polarization functions of hydrogen atom basis sets. The potential used in the subsequent scattering calculations was obtained with the aug-cc-pV6Z basis set. At large intermolecular separations, between $R = 16$ and $R = 20 \, a_0$, it was smoothly connected to the analytic long-range potential, employing a switching function[38] and the long-range coefficients ($C_6$ and $C_8$) of ref. (39). The interatomic distance of the $H_2$ molecule was set to 1.4487 $a_0$ which corresponds to the expectation value of the interatomic separation in the ground vibrational state of the hydrogen molecule. To ensure a proper convergence of the expansion into Legendre polynomials, the potential was calculated from 0 to 90 degrees with a step of 15 degrees. The projection onto Legendre polynomials was performed with the integration procedure introduced by Janssen et al.[38].

The interaction energy was corrected beyond CCSD(T) with the correction obtained as the difference between the coupled-cluster method with fourth-fold excitations (CCSDTQ) and CCSD(T) in the same basis set. Due to the enormous numerical cost, CCSDTQ calculations, which are equivalent to FCI for a four-electron system and exact for a given basis set, were performed only for linear- and T-shape geometries, with a smaller basis set, aug-cc-pVQZ and no mid-bond functions, using the CFOUR package[40] interfaced with the MRCC program[41]. The dominant source of uncertainty of the potential energy surface is incompleteness of the basis set whereas relativistic effects and the diagonal Born-Oppenheimer (DBO) correction were found to be negligible: Relativistic effects, estimated by calculating the scalar relativistic perturbation correction (mass velocity and one-electron Darwin term) are of the order of 0.01 cm$^{-1}$ in the global minimum with a well depth of 14.30 $cm^{-1}$ at 10.5 $a_0$. The DBO correction was obtained as the difference between the corrections for the dimer and infinitely separated monomers for T-shape and L-shape geometries, calculated with the aug-cc-pVQZ basis set and the Hartree-Fock wavefunction (it is not sensitive to the size of the basis set nor the quality of the wavefunction). The contribution of the DBO correction to $V_0$ is $0.044\ cm^{-1}$ in the global minimum. The basis set incompleteness affects the correlation energy only, while the Hartree-Fock interaction energy is converged to within $0.01\ cm^{-1}$ in our calculations. Truncation of the basis set results in a shallower potential and an inner turning point that is slightly shifted toward larger intermolecular distances. The basis set limit is estimated assuming a convergence pattern A+BX$^{-3}$ or A+BX$^{-3}$+CX$^{-5}$ for singlet systems and A+BX$^{-5}$ for triplet helium since the correlation energy was found to converge with the cardinal number as X$^{-3}$ for singlet electron pairs and as X$^{-5}$ for triplet pairs[42,43]. The correlation energy in the $H_2$ and $He^*$ monomers are very different in magnitude ($9031 \pm 1\ cm^{-1}$ compared to $214.7 \pm 0.01\ cm^{-1}$). For the total system, the basis set limit for the CCSD(T) calculations in the global minimum extrapolated from the aug-cc-pV6Z and aug-cc-pV7Z basis sets (in latter case we could afford for such calculations for single geometry) is $-14.62\ cm^{-1}$, whereas extrapolation from the X=5,6,7 basis set yields the limit $-14.42\ cm^{-1}$. As a conservative estimate, we take the difference between the potential obtained in aug-cc-pV6Z and the basis set limit extrapolated from X=6,7 as the uncertainty of the potential.

atomic correlation energies. *J. Chem. Phys.* **96,** 4484 (1992).